\documentclass[onecolumn]{aa}
\usepackage{graphicx}
\usepackage{txfonts}
\usepackage{graphicx}
\usepackage{natbib}
\usepackage{rotate}
\usepackage{epsf}
\begin{document}
%
\newcommand{\labeln}[1]{\label{#1}}
\newcommand{\Msolar}{M$_{\odot}$}
\newcommand{\Lsolar}{L$_{\odot}$}
\newcommand{\farcmin}{\hbox{$.\mkern-4mu^\prime$}}
\newcommand{\farcsec}{\hbox{$.\!\!^{\prime\prime}$}}
\newcommand{\kms}{\rm km\,s^{-1}}
\newcommand{\cc}{\rm cm^{-3}}
\newcommand{\Alfven}{$\rm Alfv\acute{e}n$}
\newcommand{\Vap}{V^\mathrm{P}_\mathrm{A}}
\newcommand{\Vat}{V^\mathrm{T}_\mathrm{A}}
\newcommand{\D}{\partial}
\newcommand{\DD}{\frac}
\newcommand{\TAW}{\tiny{\rm TAW}}
\newcommand{\mm }{\mathrm}
\newcommand{\Bp }{B_\mathrm{p}}
\newcommand{\Bpr }{B_\mathrm{r}}
\newcommand{\Bpz }{B_\mathrm{\theta}}
\newcommand{\Bt }{B_\mathrm{T}}
\newcommand{\Vp }{V_\mathrm{p}}
\newcommand{\Vpr }{V_\mathrm{r}}
\newcommand{\Vpz }{V_\mathrm{\theta}}
\newcommand{\Vt }{V_\mathrm{\varphi}}
\newcommand{\Ti }{T_\mathrm{i}}
\newcommand{\Te }{T_\mathrm{e}}
\newcommand{\rtr }{r_\mathrm{tr}}
\newcommand{\rbl }{r_\mathrm{BL}}
\newcommand{\rtrun }{r_\mathrm{trun}}
\newcommand{\thet }{\theta}
\newcommand{\thetd }{\theta_\mathrm{d}}
\newcommand{\thd }{\theta_d}
\newcommand{\thw }{\theta_W}
\newcommand{\beq}{\begin{equation}}
\newcommand{\eeq}{\end{equation}}
\newcommand{\ben}{\begin{enumerate}}
\newcommand{\een}{\end{enumerate}}
\newcommand{\bit}{\begin{itemize}}
\newcommand{\eit}{\end{itemize}}
\newcommand{\barr}{\begin{array}}
\newcommand{\earr}{\end{array}}
\newcommand{\eps}{\epsilon}
\newcommand{\veps}{\varepsilon}
\newcommand{\vepsdi}{{\cal E}^\mathrm{d}_\mathrm{i}}
\newcommand{\vepsde}{{\cal E}^\mathrm{d}_\mathrm{e}}
\newcommand{\ber}{\begin{array}}
\newcommand{\eer}{\end{array}}
\newcommand{\lraS}{\longmapsto}
\newcommand{\lra}{\longrightarrow}
\newcommand{\LRA}{\Longrightarrow}
\newcommand{\Equival}{\Longleftrightarrow}
\newcommand{\DRA}{\Downarrow}
\newcommand{\LLRA}{\Longleftrightarrow}
\newcommand{\diver}{\mbox{\,div}}
\newcommand{\grad}{\mbox{\,grad}}
\newcommand{\cd}{\!\cdot\!}
\newcommand{\Msun}{{\,{\cal M}_{\odot}}}
\newcommand{\Mstar}{{\,{\cal M}_{\star}}}
\newcommand{\Mdot}{{\,\dot{\cal M}}}
\title{A model for electromagnetic extraction of rotational energy \\
       and  formation of accretion-powered jets in radio galaxies}
\bigskip\bigskip

\begin{raggedright}  
 
\author{ A. Hujeirat\inst{1} \and R. Blandford\inst{2}  }  
\institute{Max-Planck-Institut f\"ur Astronomie, 69117 Heidelberg, Germany \and
Theoretical Astrophysics, Caltech, Pasadena, CA 91125, USA  }
\end{raggedright}
\offprints{A. Hujeirat, \email{hujeirat@mpia-hd.mpg.de}}

 \abstract{A self-similar solution for the 3D axi-symmetric radiative MHD equations,  which revisits
the formation and acceleration of accretion-powered jets in AGNs and microquasars, is presented.
The model relies primarily on electromagnetic extraction of rotational energy from the 
disk plasma and  forming a geometrically thin super-Keplerian layer between the
disk and the overlying corona. The outflowing plasma in this layer is  
dissipative, two-temperature, virial-hot, advective and electron-proton
dominated. The innermost part of the disk in this model is turbulent-free, sub-Keplerian
rotating and advective-dominated. This part  ceases to radiate as a standard disk, and
most of the accretion energy is converted into magnetic and kinetic energies that go into powering
the jet.  The corresponding luminosities of these turbulent-truncated disks  
are discussed.\\
In the case of a spinning black hole accreting at low accretion rates,  
the Blandford-Znajek  process is found to modify the total power of the jet, depending on the
accretion rate.   \\
     \keywords{Galaxies: jets --
            Black hole physics -- Accretion disks -- Magnetohydrodynamics
              -- Radiative transfer
               }
   }
\titlerunning{formation of accretion-powered jets in radio galaxies}
\maketitle 

\section{Introduction}
  Recent theoretical and observational efforts to uncover the mechanisms underlying jet formation 
 in AGNs and micro-quasars leave little doubt about their linkage to the accretion phenomena
   via disks \cite{Mirabel01, Livio99, Blandford01, Hujeirat03}.\\
 The innermost part of a disk surrounding a black hole 
 is most likely threaded by a large scale poloidal magnetic field 
 (-PMF) of external origin \cite{Blandford82}. Unlike standard accretion disks in which the sub-equipartition 
 magnetic field (-MF) generates turbulence that is subsequently dissipated
  and radiated away, a strong PMF in excess of thermal equipartition may suppress the generation of turbulence
  and acts mainly to convert the shear energy into magnetic energy. A part of the 
   shear-generated toroidal magnetic field (-TMF) undergoes  magnetic reconnection in the launching layer,
  whereas the other part is advected outwards to form  relativistic outflows \cite{Hujeirat03}.\\
  Indeed, the total power of the jet in M87 is approximately
 $L_\mm{J} \approx 10^{44}\,erg\,s^{-1}$ \cite{Bicknill99}.
  This is of the same order as the power resulting from a disk  accreting at the rate
  $\Mdot \approx 1.6\times 10^{-3} \Mdot_\mm{Edd}$ \cite{DiMatteo03}. Similarly, the variabilities
  associated with the microquasar GRS 1915-105, as have been classified by \cite{Belloni00},
  consists of a state "C" which corresponds to the case when the innermost part of the 
  disk diminish  \cite{Fender99}, probably a phase in which the super-luminal jet is fed with rotational and
  magnetic energies.\\ 
  Another two additional basic questions related to jet formation are:
   1) What are the contents of these jets? specifically, are they made of electron-positron $(e^{-}/e^{+})$,
      electron-proton  $(e^{-}/p)$ or a mixture of both. 2) How significant is the role of the central object 
      in the formation and acceleration of jet-plasmas?\\
   It is generally accepted that electron temperature $\vartheta_\mm{e}(\,= kT/m_\mm{e} c^2)$ of
   order unity can be easily achieved in accretion flows onto BHs, and specifically in the vicinity
   of the event horizon. At such high temperatures pair creation becomes possible.
   The efficiency of pair creation
   depends mainly on the interplay between the optical depth to scattering 
   $\tau_\mm{p} = \rho \kappa_\mm{sc} H$, on the so called "compactness" parameter  
   $\ell = \alpha_0 q^{-} H^2$ and on the strength of the magnetic field \cite{Svensson91, Esin99}. 
   H here denotes the thickness of the
   disk or a relevant length scale, $\alpha_0$ is a constant coefficient, $\rho$ is the density and
   $\kappa_\mm{sc}$ is the opacity due Thomson scattering, and $q^{-}$ is the radiative
   emissivity. Generally, the efficiency of pair creation decreases strongly with increasing the optical depth
   $\tau_\mm{p}$. It decreases also  if the strength of the magnetic field is increased. 
   Esin (1999) has studied
   pair creation in  steady hot two-temperature accretion flows for a large variety of parameters. 
   The ions here are preferentially heated by turbulent dissipation and cool via Coulomb interaction
   with the electrons, which in turn are subject to various cooling processes such as Bremsstrahlung and
   Synchrotron emission. Esin (1999) showed that two-temperature accretion flows are almost pair-free,
   whereas single-temperature accretion flows are more appropriate for pair-dominated plasmas.
   However, enhancing  thermal coupling between electrons and protons, or heating electrons
   and protons at an equal rate, and relaxing the stationarity condition may lead to 
    $e^{-}/e^{+}$-dominated plasma, provided the accretion rate is sufficiently low.\\
   In the case the BH is rotating and accreting at low rates, formation of $e^{-}/e^{+}$ jets
   though electromagnetic extraction of rotational energy from the  hole via
   Blandford-Znajek (1977) process is very likely \cite{Rees82}. Indeed, recent observations
   reveals that the jet-plasma in M87 is probably  an $e^{-}/e^{+}-$dominated jet \cite{Reynolds96}. Furthermore,
   \cite{Wardle98} reported about the detection of circularly polarized radio emission from the 
   jet in the quasar 3C 279. The circular polarization produced by Faraday conversion requires the 
   radiative energy distribution to extend down to low energy bands, 
   which indicate that $e^{-}/e^{+}-$ plasma population
   might be a significant fraction of the jet-contents.\\
   On the other hand, recently it has been argued that unless the poloidal magnetic field threading
   the event horizon is exceptionally strong, the total available power through the Blandford-Znajek process
   is dominated by the accretion power of the the disk \cite{Ghosh97, Livio99b}.
   We note, however, that the PMF can be sufficiently strong, if the innermost part of the disk collapses 
   dynamically,  while conserving poloidal magnetic flux \cite{Hujeirat03}, and which may significantly enhance the
   efficiency of the Blandford-Znajek process. \\
   In this paper we present a theoretical model for the formation of accretion-powered jets in AGNs
   and microquasars. The model relies mainly on electromagnetic extraction of rotational energy from the 
   disk plasma and  deposit it into a geometrically thin layer between the disk and the overlying
   corona (Sec. 3). In Section 4 we discuss the content of the plasma in the TL.
   The role of the Blandford-Znajek process in combination with the luminosity 
   available from the transition layer is discussed is discussed in Sec. 5.  
   We compute the total luminosity of truncated disks
   in Sec. 6, and end up with summary and conclusions in Section 7.
\begin{figure*}
\begin{center}
{\hspace*{-0.5cm}
\includegraphics*[width=6.5cm]{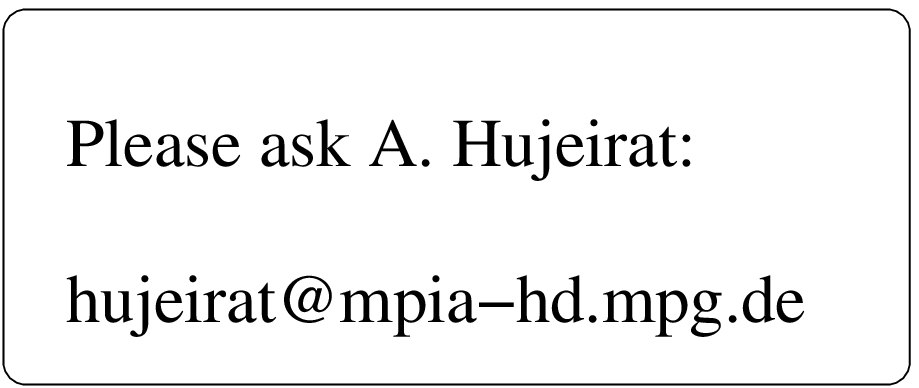}
}
\end{center}
{\vspace*{-0.4cm}}
\caption [ ] {A schematic illustration of the jet-disk connection model.
 In the outer disk,  $r\ge \rtr$, MFs are weak and advected around by fluid motions.
 Balbus-Hawley and Parker-instabilities in combination with reconnection and fast inwards 
 fluid-motions amplify the MFs and reach thermal equipartition at $\rtr$. Interior to $\rtr$,  MFs 
  become of large scale topology relative to the flow in the disk, hence adopting the profile   
  $r^{-2}$ to conserve the poloidal magnetic flux.  Here,  torsional {\Alfven} waves 
 are the dominant angular momentum carrier. They transport angular momentum from
 the disk upwards and form a super-Keplerian thin layer, where the plasma rotates super-Keplerian,
  virial-heated by reconnection, and start to centrifugal-accelerate outwards.       
  } 
\end{figure*}
\section{The model problem}
Let $r=r_\mm{tr}$ be a transition radius, where the effects of magnetic fields on
the dynamics of accretion flows become significant. This may occur if the initially
weak magnetic fields in the dissipative-dominated disk-plasma are amplified via dynamo
action, in which Balbus-Hawley (1991) instability in combination with the Parker instability 
results in a topological change of the magnetic field
from a locally disordered into well-ordered large scale magnetic field (see Fig. 1). 
At this radius the MFs are in equipartition with the thermal energy of the plasma, i.e., 
$\beta = E_\mm{mag}/E_\mm{th} = 1.$

Interior to  $r_\mm{tr}$  the poloidal magnetic field $\Bp$ is predominantly of large
scale, and magnetic flux conservation implies that PMF 
increases inwards obeying the power 
law\footnote{$\Phi = 2 \pi r^2 B_\mathrm{p} \approx 2 \pi r^2 
B_\mathrm{\theta} = const.$}
 $\Bp \sim r^{-2}.$ 
Such a strong  PMF suppresses the generation of turbulence,
 whenever $\beta$ exceeds unity.
In this case, the heating mechanisms that hold the plasma  against runaway cooling are
magnetic reconnection, adiabatic compression and other non-local sources such as 
radiative reflection,  Comptonization and conduction of heat flux from the surrounding hot media. \\
Given a plasma threaded by an $r^{-2}-$ PMF, the angular velocity
must deviate significantly from its corresponding Keplerian profile. This is because  torsional {\Alfven} waves (-TAWs) 
extract rotational energy from the disk plasma on the time scale: $\tau_\mm{TAW} = H_\mm{d}/\Vap,$
which decreases strongly inwards. $\Vap$ here is the {\Alfven} speed due to the PMF.
The TAW crossing time of the disk, i.e.,  $\tau_\mm{TAW}$,  is of the same order, or it can be even shorter
than the dynamical time scale
$\tau_\mm{dyn} = r/\Vt$. In this case the angular velocity becomes sub-Keplerian, 
and the disk becomes pre-dominantly advection-dominated.
If  $\tau_\mm{TAW} < \tau_\mm{dyn},$ accretion may terminate completely.\\ 
TAWs, which carry with it rotational energy from the disk-plasma, 
may succeed to propagate through the overlying corona without being dissipated. This corresponds
to magnetic braking of the disk, in which the rotational energy is transported from
the disk and deposited into the far ISM. However, since the PMF-lines rotate faster as
the equator is approached, the PMF  winds up, intersect and subsequently reconnect, and so 
inevitably terminate the propagation of TAW into higher latitudes. Alternatively,
the vertical propagation of the TWAs may terminate  via magnetic reconnection at the surface
of the disk, establishing thereby a highly dissipative transition layer (TL), i.e., chromosphere,
where the toroidal magnetic fields intersect, reconnect and subsequently
heat  and particle-accelerate the plasma in the TL. 
The latter possibility is more plausible, mainly because:
\ben
\item The vertical profile of the angular velocity generate a toroidal magnetic field
of opposite signs, i.e., anti-parallel toroidal flux tubes, which give rise to magnetic 
 reconnection of the toroidal magnetic field lines (-TMF). The
reconnection process terminates the vertical propagation of the  TAWs, inducing thereby a magnetic
trapping of the rotational energy in the TL.  As a consequence, a centrifugally  induced outflow
is initiated, which carries with it rotational, thermal and magnetic energies.
\item  Unlike stellar coronae that are heated from below, BH-coronae have been found
to be dynamically unstable to heat conduction \cite{Hujeirat02}. The effect of conduction is 
to transport heat from the innermost hot layers into the outer cool envelopes. Furthermore,
 noting that the Lorenz and centrifugal forces in an  axi-symmetric flow 
attain minimum in the vicinity of the  rotational axis, we conclude
that this region is inappropriate for initiating outflows.
\item  In a stable flow-configuration, an isolated Keplerian-rotating particles in the disk
       region move along trajectories with minimum total energy. If these particles
       are forced to move  to higher latitudes while conserving their angular momentum, 
       then their rotation velocity  becomes super-Keplerian, and therefore start to
      accelerate outwards as they emerge from the disk surface. Since the speed of
these motions and the associated strength of the generated toroidal flux tubes increase inwards, 
 the magnetic flux tubes are likely to interact and subsequently to reconnect. 
\item Observations reveals that the radio luminosity in the vicinity of the nucleus of various AGNs
      with jets acquire a significant fraction of the total luminosity.  This 
     indicates the necessity
     for an efficient heating mechanism that allow the electrons
     to continuously emit Synchrotron radiation on the {\Alfven} wave crossing time.
     Noting that $\Bp \sim r^{-2}$,  and that
     the plasma in the innermost part of the disk is turbulent-free, 
     we propose that magnetic reconnection of the TMF is most reasonable
     mechanism for heating the plasma in the TL. This agrees with the proposal of   
     \cite{Ogilvie01} who argued that jet launching requires thermal assistance.
\een
\subsection{The velocity field}
The problem we are addressing in this paper is: assume that the innermost part of
the disk is threaded by $r^{-2}-$ PMF. What are the most reasonable
power-law distributions of the other variables in both, the disk and in the TL, that give
rise to inflow-outflow configuration, and which simultaneously satisfies the set of
the steady 3D axi-symmetric radiative MHD equations. As we shall see in the next sections, 
there are several signatures that hint to $\Omega  \sim r^{-5/4}$  as the optimal
profile (see Fig. 2). Nevertheless, we will consider this profile  as an assumption.\\
Therefore, in the disk region the following profiles are assumed:
\beq
   \Omega(r)  =  \Omega_0 X^{-5/4}, \,
   \Bp (r)     =   \Bpz(r)  \approx B_0 X^{-2}, 
\eeq
 where  $X= r/\rtr$. \\ 
To assure a smooth matching of the variables interior to
$\rtr$ with those of the standard disk exterior to $\rtr$, we set 
 $\Omega^2_0 = GM/{\rtr^{3}}$,   and
 $ {B_0^2}= {B^2}(r=\rtr) 
   =(12 \pi {\cal{R}_\mm{g}}/\mu (\gamma -1)) \rho T|_{r=\rtr}$,
 Inserting  $\Omega(r)$ in Equation A.2 (see Appendix), and taking into
account that the internal  and magnetic energies of the plasma at 
$\rtr$ are negligibly small compared to the gravitational energy of the
flow, the radial velocity then reads:
\beq
 U_\mm{d}(r) = [U^2_{0} + \DD{2}{\eps^2} \DD{1}{\rtr} (\DD{1}{X} - 1)]^{1/2}. 
\eeq
As in classical disks, we set $U_{0} = U(r=\rtr) = - (2/3) \nu/\rtr$,
and  $\nu = \alpha_\mm{ss} H_\mm{d} V_\mm{s}$ for describing turbulent 
viscosity. \\
\begin{figure*}
\begin{center}
{\hspace*{-0.5cm}
\includegraphics*[width=12.5cm, bb=50 395 515 725,clip]{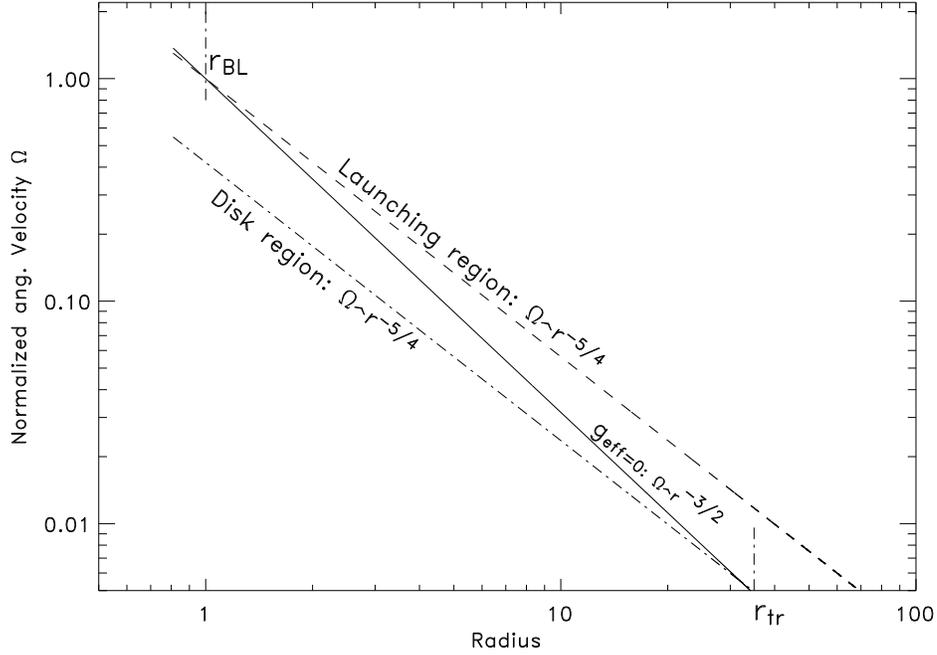}
}
\end{center}
{\vspace*{-0.4cm}}
\caption [ ] {The profiles of the angular velocity in the innermost part of
              the disk and in the launching region.
              TAWs extract rotational energy from the disk-plasma efficiently
              and deposit it into the plasma adjusting to the disk-surface. As a consequence,
              the inflow adopts the sub-Keplerian profile  $\Omega \sim r^{-5/4}$ in the disk-region,
              while the outflow adopts the super-Keplerian profile $\Omega \sim r^{-5/4}$ in the 
              TL. Note that the up-stream boundary conditions determine whether the flow is sub-
              or super-Keplerian rotating.     
  } 
\end{figure*}
The radial accretion rate in the disk region is obtained by $\theta-$integrating the 
continuity (Eq. A.1):
\beq
  \DD{\D}{\D r} \Mdot^\mm{d}_\mm{r} = 
       2 r \cos{\theta_\mm{d}} (\rho V)|_{\theta_\mm{d}},
\eeq
where $\Mdot_\mm{d}^\mm{r} = r\Sigma_\mm{d} U_\mm{d}$ is the radial accretion rate in the disk region, 
and $\Sigma_\mm{d} = \int^{\theta_\mm{d}}_{-\theta_\mm{d}}{r\cos{\theta} \rho_\mm{d} d\theta} $.
$\theta_\mm{d}$ is the angle such that $\tan{\theta_\mm{d}} = H_\mm{d}/r,$
where $H_\mm{d}$ is the vertical scale height of the disk (see Fig. 3). 

Furthermore, it is reasonable to assume that the radial dependence of the 
density at the surface of the disk does not differ significantly from the density at the
equator. Thus, $\rho(r,\theta=\theta_\mm{d})/ \rho(r,\theta=0) = const. << 1$.\\ 
\begin{figure}[htb]
\begin{center}
{\hspace*{-0.5cm}
\includegraphics*[width=7.5cm, bb=0 0 368 262,clip]{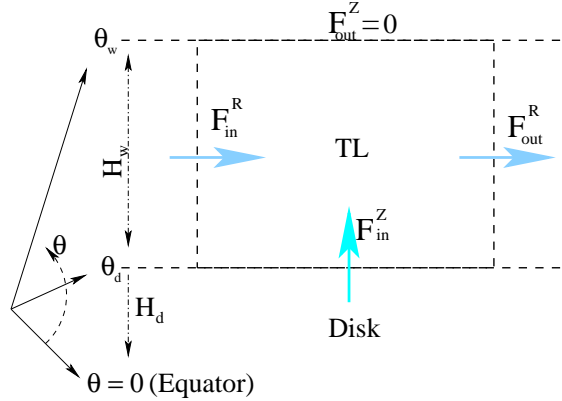}
}
\end{center}
{\vspace*{-0.4cm}}
\caption [ ] {The material fluxes governing an arbitrary volume cell 
              in the TL   are shown. $\theta_\mm{W,d}$ 
     are latitudes that correspond to the locations of the interface between the disk
   and the TL, and to the upper surface of the TL, respectively.
     ${\rm F}^\mm{Z}_\mm{in,out}$ and ${\rm F}^\mm{R}_\mm{in,out}$  denote the
      material fluxes across the vertical and radial surfaces. The width of the cell 
     is $H_\mm{W}$, and the vertical material flux across $\theta_\mm{W}$ is set to vanish.
  } 
\end{figure}
Similar to the disk region, the following angular velocity and the PMF profiles are
set to govern the plasma in the launching region:
\beq
   \Omega(Y)  =  \Omega^\mm{W}_0 Y^{-5/4},\,
   \Bp(Y)  \approx \Bpr    =  B^\mm{W}_0 Y^{-2}, 
\eeq
 where  $Y= r/\rbl$, and $\rbl$ is the innermost radius of the TL, where the
 effective gravity is set to vanish. This implies that at $Y=1$, the angular velocity
 approaches the Keplerian values: 
 $\Omega^\mm{W}_0 = \Omega^\mm{W} (Y=1) = \sqrt{GM/{\rbl^{3}}}$. 
 Inspection of Eq. A.2 implies that the radial velocity in the 
super-Keplerian region has the profile:
 \beq
    U_\mm{W}(Y)      =  [U^2_\mm{W,0} + \DD{2}{\eps^2} 
    \DD{1}{\sqrt{\rtr}} (1 - \DD{1}{\sqrt{Y}}]^{1/2}.
\eeq
where $U_\mm{W,0}$ is the radial velocity at $\rbl$. However, since the effective 
gravity has a saddle point at  $\rbl$, the radial velocity must vanish.
 Interior to  $Y=1$,
 the plasma is gravitationally bound, so that the plasma and the associated
 rotational energy will be advected inwards and disappear in the hole. In the case of a spinning BH,
 $\rbl$ is likely to be located inside the ergosphere.\\
The outward-oriented material flux in the TL is:
\beq
  \Mdot^\mm{W}_\mm{r} = r\Sigma^\mm{W} U^\mm{W},
\eeq
where $\Sigma^\mm{W}  = \int^{\theta_\mm{W}}_{\theta_\mm{d}}{r\cos{\theta} \rho^\mm{W}  d\theta} $.
$\theta_\mm{W}$ is the angle such that $\tan{\theta_\mm{W}} - \tan{\theta_\mm{d}}= H_\mm{W}/r,$
and $H_\mm{W}$ is the vertical scale height of the launching layer. \\
Global mass conservation of the plasma both in the disk and in TL regions
requires that:
\beq
\DD{\D}{\D r} \Mdot^\mm{d}_\mm{r} = 2 \DD{\D}{\D r} \Mdot^\mm{W}_\mm{r},
\eeq
where mass loss from both sides of the disk has been taken into account, and
where have assumed a vanishing vertical flux of matter across $\theta_\mm{W}$ (see Fig. 2 ). \\ 
\subsection{ The geometrical thickness of the launching region }
 
Since the plasma in the TL rotates super-Keplerian, the horizontal 
component of the centrifugal force, i.e.,
$({\tan{\theta}}/{\eps^2})({\Vt^2}/{r})$ (see Eq. A.3),  
acts to compress the plasma in the launching region toward the equator and its  
collapse. Inspection of Eq. A.3 reveals that there are four forces that 
 may oppose collapse: gas and turbulent pressures,  poloidal and 
 toroidal magnetic fields. In terms of Eq. A.3, the dominant terms required for
 establishing such an equilibria are: 

\beq
\DD{\tan{\theta}}{\eps^2}\DD{\Vt^2}{r} = \left \{ \begin{array}{lcl}
 \DD{1}{\rho} \DD{1}{r} \DD{\D}{\D \theta}P_\mm{gas}: &  &  {\rm ion\,pressure} \,\,(I)\\
 \DD{1}{\rho} \DD{1}{r} \DD{\D}{\D \theta}P_\mm{tur}: &  &  {\rm turbulent\,pressure} \,\,(II)\\
 \DD{1}{\rho} \DD{1}{r} \DD{\D}{\D \theta}\Bp^2   :   &  &  {\rm poloidal\,MFs} \,\,(III)\\
 \DD{1}{\rho} \DD{1}{r} \DD{\D  }{\D \theta} \Bt^2 :   &  &  {\rm toroidal\,MFs} \,\,(IV).
 \end{array} \right. 
\eeq
Equivalently, the relative geometrical thickness of the TL reads:  
\beq
\DD{H_\mm{W}}{r} = \left \{\begin{array}{lcl}
   \eps \times ({V_\mm{S}}/{\Vt}):&      &\, \rm{ion\,pressure } \,\,(I)\\
   \eps \times ({V_\mm{tur}}/{\Vt}):&      &\, \rm{turbulent\,pressure } \,\,(II)\\
   \eps \times ({V^\mm{P}_\mm{A}}/{\Vt}):&     &\, \rm{poloidal\,MFs } \,\,(III)\\
   \eps \times ({V^\mm{T}_\mm{A}}/{\Vt}): &   &\, \rm{toroidal ,MFs} \,\,(IV),
 \end{array} \right. 
\eeq
were we have excluded magnetic tension from our consideration, because the 
plasma in the TL is dissipative.
Global magnetic flux conservation, i.e., $\nabla \cdot B=0$, requires that $\Bp$ must bend
as it emerges from the disk region, provided that $\Bp$ is of large scale topology
and is advected inwards with the matter. In this case,
the above-mentioned third possibility, i.e., case III,  applies to those MFs in which $\Bpr$ decreases 
vertically and therefore  may hold the plasma in the TL against vertical collapse.
 This means that the MF-lines in the TL must point toward the central BH as they emerge 
from the disk (see Fig. 4). Although such configurations are not in view with our classical
expectation about magnetic-induced launching, the radiative MHD
calculations do not appear to exclude such solutions, provided that TL-plasma is highly dissipative. 
On the other hand, if the poloidal MF-lines 
in the TL point away  as they emerge from the disk, which is the case we are considering in this study, 
 $\Bpr$ must increase
in the vertical direction, and therefore may enhance the collapse. Therefore, taking into
account that the electron thermal energy  and the PMF-energy are relatively small compared to the 
rotational energy in the super-Keplerian layer, and that $\Bt-$profile has turning points
in the TL, we may conclude that $P_\mm{tur}$ is the most reasonable force to oppose  the TL-collapse. \\
To elaborate this point we note that when TAWs start propagating from the disk upwards while carrying 
angular momentum, a profile such as shown in Fig. 5 would result. This profile has $\nabla \Omega$ 
of mixed signs, which therefore induces the formation of toroidal flux tubes whose the associated 
 electric currents $\vec{j}$ move anti-parallel around
the axis of rotation, and which gives rise to partial magnetic reconnection (Fig. 5).\\
 The strong shear in the TL is capable to produce magnetic
energy that is in equipartition with the rotational energy, i.e., $V^\mm{T}_\mm{A} \approx \Vt$,
which is necessary to maintain the steadiness stationarity
of the solution of the $\Bt-\Vt$ coupled system. \\
We note that magnetic reconnection is a common phenomena  in astrophysical MHD-flows, 
which is generally associated with changes in the magnetic field topology and partial loss of the magnetic 
flux. 
In the solar case, for example,  magnetic reconnection
is one of the main mechanisms underlying the eruption phenomena of the solar flares. Approximately
$10^{32}$ ergs are liberated in each event and on the time scale of few minutes. Although
the sun is not as compact  as old stars such as a white dwarf or a neutron star,  solar flares appear to be associated with $\gamma-$ray emission,
electron-positron annihilation at 511 Kev, and most impressively, with the neutron capture
by the hydrogen nucleus which occurs at about two mega electron volts, revealing thereby that 
acceleration of particles up to relativistic velocities is an important ingredient of the reconnection process.\\
Yet, the TL is located deep in the gravitational well of the central BH, and if reconnection
occurs, then the energetics associated must be much powerful than in the solar flares,
and a significant   populations of $e^-/e^+$   are to be expected in the TL. \\
In plasma physics, reconnection occurs normally on microscopic scales.
In the present study  we assume that reconnection can be described by the magnetic diffusivity:
\beq
 \nu_\mm{mag} = H_\mm{W} V_\mm{tur},
\eeq 
 where $H_\mm{W}$ is the width of the TL and $V_\mm{tur}$ 
is  the velocity that is associated with magnetic reconnection. We adopt the Petscheck-scenario for describing the
the reconnection velocity \cite{Priest94}:
\beq
  V_\mm{rec} = V_\mm{tur} = \alpha_\mm{mag} \Vat  /\log{Re},
\eeq
where  $\alpha_\mm{mag} $ is a constant less than unity and $\log(Re)$ is the logarithm of the magnetic Reynolds 
number.\\
Inserting $V_\mm{tur} =  \alpha_\mm{mag} \Vat  /\log(Re)$ in Eq. 9/II,  
we obtain\footnote{Without considering $\eps$; it will be determined later.}:
\beq
   \DD{H_\mm{W}}{r} = \DD{V_\mm{tur}}{\Vt} = \DD{\alpha_\mm{mag}}{\log(Re)}
\eeq
From the definition of the magnetic Reynolds number:
\beq
Re = \DD{\rm{Inertial\,force}}{\rm{Viscous\,force}} = (\DD{Bp}{\Bt}) (\DD{\Vt}{V_\mm{tur}}) 
= (\DD{Bp}{\Bt}) (\DD{\log{Re}}{\alpha_\mm{mag}}),
\eeq
we obtain an equation for the magnetic Reynolds number:
  
\beq
 \log(Re) - [\alpha_\mm{mag} (\DD{\Bt}{\Bp})] Re = 0.
\eeq
Obviously, $\alpha_\mm{mag}$ must be relatively small for this equation to provide a 
positive $Re$ (see Fig. 6), which means that the TL must be geometrically thin, and that
 $\eps \approx \alpha_\mm{mag}/\log{Re}  = H_\mm{W}/r$.
In order to assure that the outflow leaving the system has sufficient   
TMF to collimate and reach large Lorentz factors, we require that  only
a small fraction of the total generated 
TMF dissipate in the TL through reconnection events, while the rest is
advected outwards. Therefore, the advection time scale should be equal to the amplification time
scale of the TMF, i.e., $\tau_\mm{adv} = \tau_\mm{amp}$. This yields:
$\Bp/\Bt = H_\mm{W}/r = \eps$.   \\
\subsection{The profile of the toroidal MF}
In the disk region, where the plasma is turbulence-free, TAWs are the dominant
angular momentum carrier.
Assuming the plasma to be instantly incompressible, 
the equation describing the propagation of the magnetic torsional
wave  reads approximately: 

\beq
 \DD{\D^2}{\D^2 t}B_\mathrm{T}  \cong {V^2_\mathrm{A}}   
       \Delta {B_{\rm T}}.
\eeq
$\Delta$ denotes the two-dimensional Poisson operator in spherical geometry, and 
$V_\mm{P}^\mathrm{A}$ ($\doteq B_\mathrm{p}/\sqrt{\rho}$)  is the {\Alfven} speed.
The action of these waves is to
extract angular 
momentum from the disk to higher latitudes (see Fig. 4). Note that $B_\mathrm{p}$
 determines uniquely the speed of propagation, hence the  efficiency-dependence  
 of angular momentum transport on the $B_\mathrm{p}-$topology.
These torsional {\Alfven} waves (-TAWs) propagate in the vertical direction on the time scale:
\[ \tau_{\TAW} = \DD{H_\mm{d}}{V_\mathrm{A}} = \DD{H_\mm{d}\sqrt{\rho_\mm{d}}}{\Bpz}.  \]
Obviously, since $H_\mm{d}$ attains a minimum value and $V_\mathrm{A}$ attains
a maximum value at the innermost boundary, $\tau_{\TAW}$ can be extremely short so that, in the absence
of flux losses, 
a complete magnetic-induced termination of accretion  should not be excluded.
 \\ 
To maintain dynamical stability of the disk and teh TL, extraction of angular momentum should be compensated
by radial advection from the outer adjusting layers, i.e., 
\[\DD{1}{r^2}\DD{\D}{\D r} (r^2 U \ell) = \Bpz \DD{\D \Bt}{\D \theta}, \]
where $\ell = r^2 \cos{\theta}^2 \rho \Omega $ is the angular momentum. \\
When $\theta-$integrating
this equation from $-\thetd$ to $\thetd$, we obtain:
\beq
\DD{1}{r^2} \DD{\D}{\D r}(\Mdot^\mm{r}_\mm{d} r^2 \Omega) = \Bpz \Bt,
\eeq
where $\Bt$ corresponds to the values in the TL, and which
 can be determined by matching both the disk- and the TL-solutions.  \\
Taking into account that $\Mdot^\mm{d}_\mm{r}$ varies slowly with radius, 
and inserting $ \Omega \sim r^{-5/4}$ and $\Bp\sim r^{-2}$, we obtain $\Bt \sim r^{-1/4}$,
which   depends weakly on the radius (see Fig. 9). \\ 

\begin{figure}[htb]
\begin{center}
{\hspace*{-0.5cm}
\includegraphics[angle=-0,width=6.5cm]{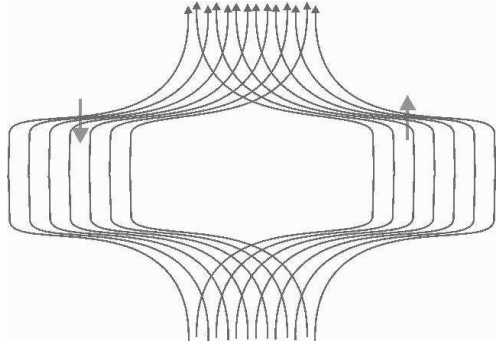}\\
}
\end{center}
{\vspace*{-0.4cm}}
\caption [ ] {Possible configurations of converging and diverging poloidal magnetic
                     field in the disk and the TL. The left configuration
                 corresponds to the case in which the magnetic pressure at the interface
                 between the disk and the TL increases vertically, and therefore enhances
                  the collapse of the TL. The right case corresponds to magnetic pressure
                  that opposes collapse. 
  } 
\end{figure}
\subsection{The profiles of the other variables}
\ben
  \item To have stationary solutions, the energy due to the toroidal magnetic field 
    in the TL must be is in equipartition with the rotational energy, i.e., $E^\mm{T}_\mm{mag}=E_\mm{rot}$.
   This means that:
\[ V^\mm{T}_\mm{A} = \DD{\Bt}{\sqrt{\rho}} = \Vt, \LRA \rho = (\DD{\Bt}{\Vt})^2 = const. \doteq \rho^0_\mm{W}.\]
Thus the disk must supply the plasma in  the TL with the optimal material flux across the interface, 
   so to maintain the density constant.\\
When r-integrating the equation of mass conservation, Eq. 7, we obtain: 
\beq
 \Mdot^\mm{d}_\mm{r} =  \Mdot^\mm{d}_\mm{r=\rtr} 
    + 2\,(\Mdot^\mm{W}_\mm{r} - \Mdot^\mm{W}_\mm{r=\rtr}) = \Mdot^\mm{d}_\mm{r=\rtr} + 2\Delta \Mdot^\mm{W}.
\eeq
 This gives the density profile along the equator:  
\beq
\rho_\mm{d} = \DD{\Mdot^\mm{d}_\mm{r}}{r H_\mm{d} U_\mm{d}} = \DD{\Mdot^\mm{d}_\mm{r}}{\eps r^2 U_\mm{d}},
\eeq
which is a function of radius and of $\rho^0_\mm{W}$.\\
The density in the TL then reads:
\beq
\rho_\mm{W,0} = (\Mdot^\mm{d}_\mm{\rtr} - \Mdot^\mm{d}_\mm{r=rin}) r^{-7/4}{\cal F}^{-1}/2,,\,\,
{\rm where} \,\,{\cal F} = [\sqrt{r}-1]^{1/2}
\eeq
Although $\rho_\mm{W,0}$ can be treated as an input parameter, we provide here its  value
relative to the central density at the equator. Taking into account that $U_\mm{W}$ vanishes at
 $\rbl$ and applying mass conservation to an adjusting volume cell in the TL, we obtain the  rough 
estimate:
\beq
 \rho_\mm{W,0} =  [\rho_\mm{\theta_d}] [(\DD{V_\mm{\theta_d}}{U^\mm{W}})][(\DD{r}{H_\mm{W}}] 
   = [\eps \rho_\mm{c}][\eps^2][1/\eps] = \eps^2 \rho_\mm{d}(r=\rbl). 
\eeq
In deriving this result, we have used $V_{\theta_d}=\eps^2 U_\mm{d} $, which relies on asymptotic expansion 
of the variables $q= q_0 + \eps q_1 + \eps^2 q_2 + ...$, where $q_0$ is the value corresponding to an 
 equilibrium  state, i.e., hydrostatic  equilibrium in the vertical direction 
\cite[][see the references therein]{Regev}.
Thus, replacing
${\D \Mdot^\mm{W}_\mm{r}}/{\D r}$ by $ {\Mdot^\mm{W}_\mm{r}}/{r}$ in Eq. 7, we may obtain
a more accurate value for the vertical velocity across the interface at $\theta_\mm{d}$: 
\beq
 V(r,\theta_\mm{d}) = \DD{\rho^\mm{W}_0 r^{5/4} {\cal F}}{
    [ \Mdot^\mm{d}_\mm{0} - 2 \rho^\mm{W}_0 (\Mdot^\mm{W}_\mm{0} + r^{7/4}  {\cal F})]}.
\eeq
The outward-oriented material flux compared to the accretion rate is displayed in Fig.7.
The corresponding 2D profiles of the poloidal and toroidal components of the velocity field 
are shown in Fig. 8.
\begin{figure}[htb]
\begin{center}
{\hspace*{-0.5cm}
\includegraphics*[width=6.5cm, bb=22 30 1029 610,clip]{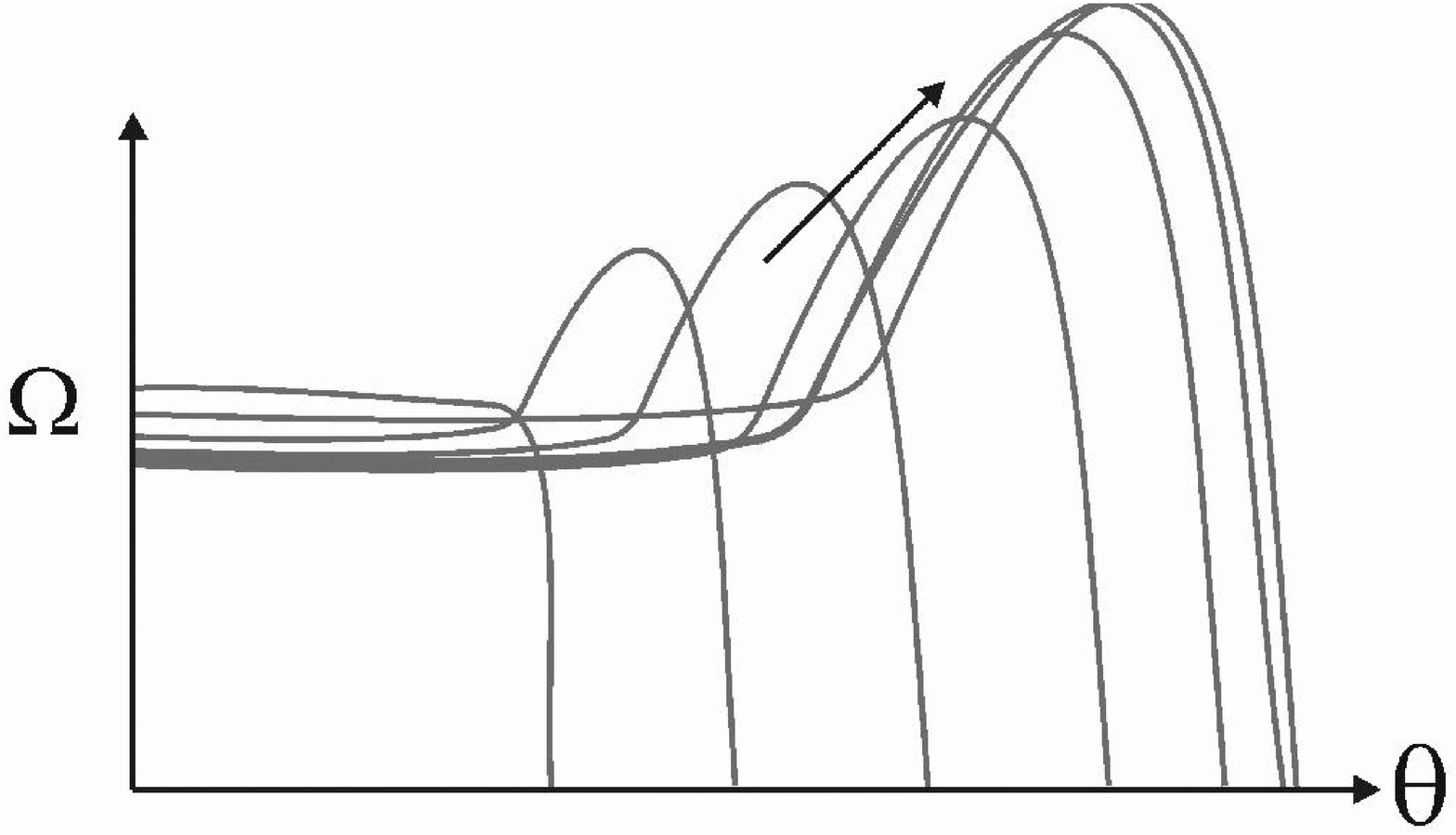} \\
\includegraphics*[width=6.5cm, bb=0 0 462 324,clip]{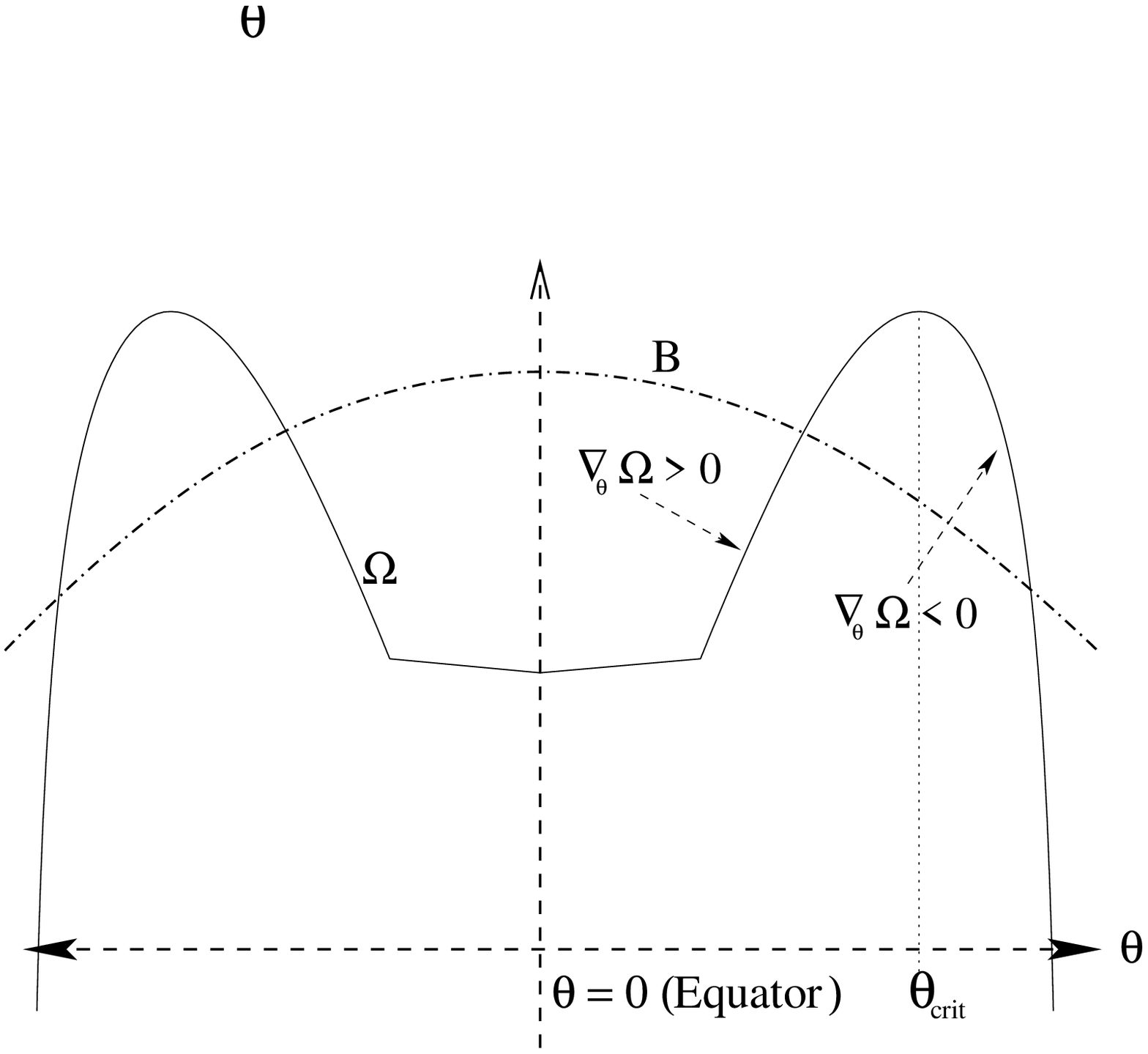}\\
\includegraphics*[width=6.5cm]{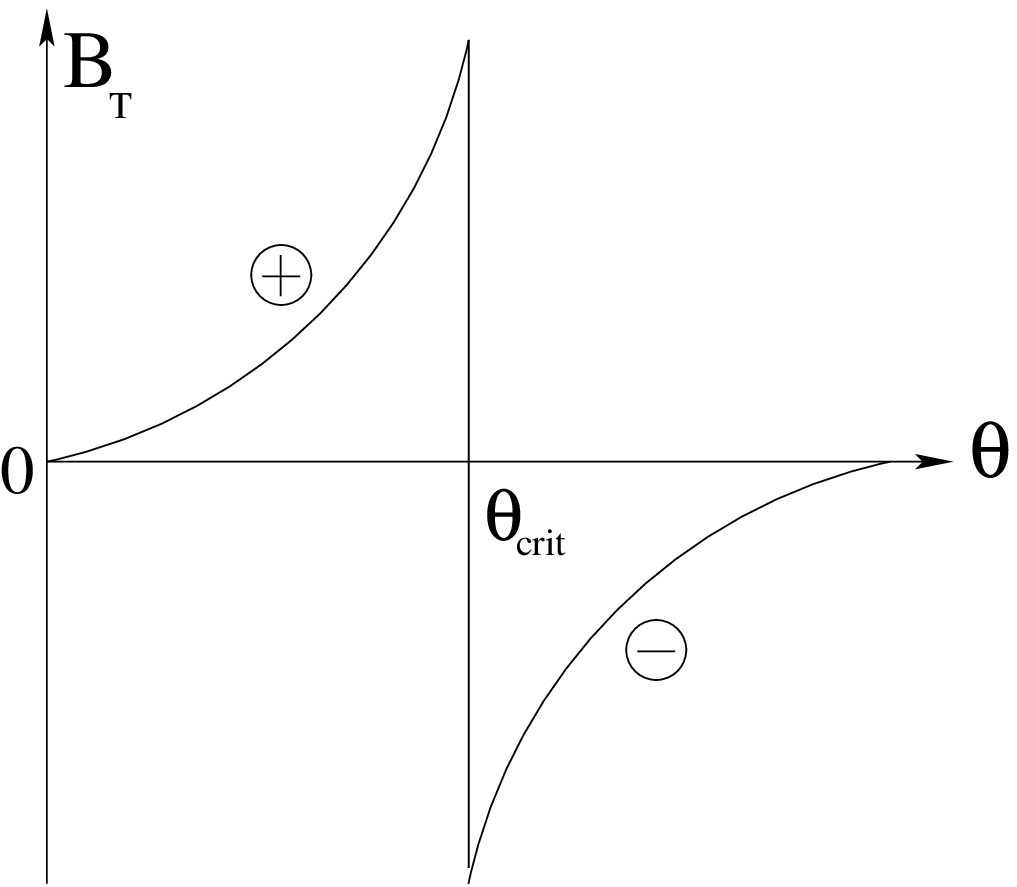}
}
\end{center}
{\vspace*{-0.4cm}}
\caption [ ] {A schematic description for a time-sequence of the angular velocity profile 
                    in the vertical direction. TAWs transport angular momentum vertically, enforcing the matter
         in the disk to rotate sub-Keplerian, and  super-Keplerian in the TL (top).
          The resulting $\Omega-$profiles acquire negative and positive spatial derivatives (middle), which in the
               presence of a large scale PMF, generate toroidal flux tubes  of opposite
                signs (bottom) that subsequently reconnect and annihilate.
  } 
\end{figure}
\item  The profile of the magnetic diffusivity in the TL reads: $\nu_\mm{mag} = H_\mm{W} V^\mm{T}_\mm{A} = \eps r \Bt/\sqrt{\rho} \sim r^{3/4}.$
\item The ion temperature is obtained by requiring that the advection  
        and heating time scales to be equal, i.e., $\tau_\mm{adv} = \tau_\mm{heat}$, where
        $\tau_\mm{heat} = C_\mm{V} \rho T_\mm{i}/\Phi$. $\Phi$ is the dissipation function:
        $\nu_\mm{mag} |\nabla\times B|^2 \approx \nu_\mm{mag} \Bt^2/H^2_\mm{W}$. This gives
        $T_\mm{i} \sim r^{-1/2}$, or more accurately, 
\beq
       T_\mm{i} = T^0_\mm{i} Y^{-1/2}$, where $T^0_\mm{i} = 1/\eps^2.
\eeq
  \item The electron temperature is extremely sensitive to the strength of the magnetic field. Since 
         the heat liberated through  magnetic reconnection can heat both the electrons and protons
         equally, the temperature of the electrons can be found  by equalizing the Synchrotron
         cooling rate to the heating rate. In order to calculate the Synchrotron cooling,
      we must first find the critical frequency, below which the media becomes optically thick
     to Synchrotron radiation, i.e, $\nu_\mm{c}$, below which the emission follow the 
     Raleigh-Jeans blackbody emissivity profile. This requires solving the equation:
\beq
      \int{\varepsilon^{\nu}}\,d Vol = \int{B_{\nu}} d \,S,
\eeq
where $\varepsilon^{\nu}$ and $B_{\nu}$ are the Synchrotron and Raleigh-Jeans blackbody emissivities,
respectively.   Having calculated $\nu_\mm{c}$, the Synchrotron cooling $\Lambda_\mm{syn}$ 
is then calculated by dividing the total luminosity from the surface of the TL divided by 
its corresponding volume-integrated frequency from $\nu =0$ up to  $\nu_\mm{c}$, where the medium is
assumed to be self-absorbed for $\nu \le \nu_\mm{c}$. This gives \cite{Esin96}:

\beq
 \Lambda_\mm{syn} \approx \DD{2\pi k T_e \nu^3_\mm{c}}{3 H_\mm{W} c^2}.
\eeq
Synchrotron emission at higher frequencies is assumed to be negligibly small. 
In terms of the density and magnetic fields, the following approximation due to 
  can be used \cite{Shapiro83}:\\
\( \Lambda_\mm{syn} = 2.16\times 10^4 B^2_3 T^2_9 \rho_{10},\)
where $B_3 = B/10^3 G$, $T_9 = T/10^9 K$ and $ \rho_{10}= \rho/10^{-10} gr.$\\
   Therefore, from the equalizing the heating to cooling rate, we obtain the following profile
 for the electron temperature: $ T_\mm{e} = T^0_\mm{e}\, Y^{-5/8},$ and
            $T^0_\mm{e} =1/\eps.$
\item  It should be stressed here that only a small fraction of the total generated TMF is allowed to undergo
       magnetic reconnection in the TL. This is an essential requirement
       for not obtaining radio luminosity that dominate the total power emerging from
      jet-bases, and to assure that the outflow is associated with TMF-energy sufficient enough 
      for collimating the  outflows into jets.  
     Thus, comparing  the rate of heating via magnetic reconnection  with the 
     generation rate of the TMF, we obtain: 
\beq
     \DD{(1/\tau_\mm{Diss})}{(1/\tau_\mm{gen})} = \DD{\tau_\mm{gen}}{\tau_\mm{Diss}} =
     \eps (\DD{\Bt}{\Bp}) (\DD{\nu_\mm{mag}}{H_\mm{W} \Vt}) = \DD{\sqrt{2}\alpha_\mm{mag}}{\log(Re)} \approx \eps,
\eeq 
 where we have used $\nu_\mm{mag} =  H_\mm{W}  V^{T}_\mm{A}$ and 
$\Bp/\Bt = H_\mm{W}/r = \eps$.  
 Consequently, an $\eps-$fraction of the total generated toroidal magnetic energy undergoes magnetic
 reconnection, while the rest is advected with  the relativistic outflow.  
\een
We now turn to find out the appropriate profiles of the electron- and proton-temperatures
in the disk region. 
\begin{figure}[htb]
\begin{center}
{\hspace*{-0.5cm}
\includegraphics*[width=6.5cm]{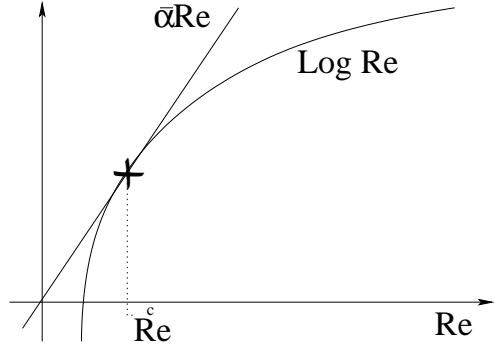}
}
\end{center}
{\vspace*{-0.4cm}}
\caption [ ] {The profiles of $\bar{\alpha}Re$ and $\log{Re}$ versus the magnetic Reynolds
              number $Re$, where   $\bar{\alpha}= \alpha_\mm{mag} (\DD{\Bt}{\Bp})$.
               The intersection
               point $Re^\mm{c}$ corresponds to the solution of Equation 14.  
               } 
\end{figure}
 The electrons in the disk region are heated mainly by adiabatic compression,
conduction and by other non-local energy sources. However, they may suffer of an extensive 
cooling due Synchrotron emission. We assume that at the interface between the 
disk and the TL, there is a conductive flux that is sufficiently large so to compensate the cooling
of the disk-electrons through Synchrotron emission. In other wards, we require that $\tau_\mm{Syn} = \tau_\mm{heat} $. In this case
the electron temperature adopts the adiabatic profile; $T_\mm{e} \sim r^{-1}.$ 

 How does the ion-temperature correlate with that of the electrons in the disk region?\\
Assume that the electrons and the  protons to have Maxwell velocity distributions. The time scale
required to establish an equilibrium \cite{Spitzer56} is:
\beq
 \tau_\mm{therm} = 4.11 (T_9^{3/2}/n_{14}) (20/ln{\Lambda})\,s,
\eeq
where $T_9 = T/10^9\,K$, and the proton number density is given as $n_{14}=n/10^{14}$.
Therefore, in the disk region $\tau_\mm{therm} $ is of the same order as or greater than the hydrodynamical
time scale if $n\ge 10^{10}$, which implies that $ T_\mm{e} \approx T_{p}$ for reasonable accretion rates. In the transition layer, however,
$n(TL)\sim \eps^2 n(disk)$. Therefore $ T_\mm{e} $ and $T_{p}$ can be significantly  
different.  
\begin{figure}[htb]
\begin{center}
{\hspace*{-0.5cm}
\includegraphics*[height=7.0cm,width=6.5cm,bb=40 150 275 475,clip]{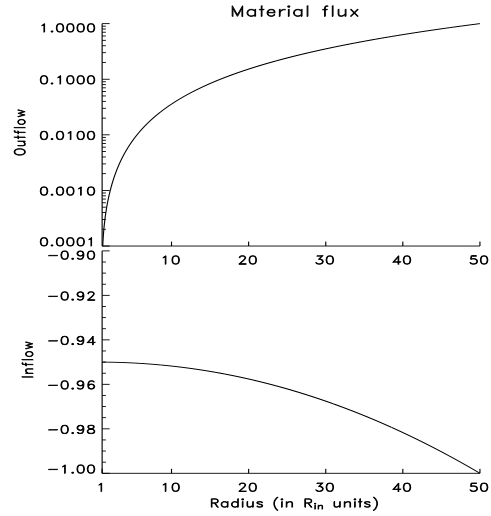}
}
\end{center}
{\vspace*{-0.4cm}}
\caption [ ] { The radial profiles of the accretion rate in the disk region, and the 
                outwards-oriented material flux in the launching layer, $F_\mm{outflow},$
                which applies for  $r \le \rtr$. 
                Note that  $F_\mm{outflow}$ vanishes at the inner boundary $\rbl$, where
                the effective gravity is set to vanish.
               Assuming $\eps=0.1=H_\mm{d}/r$, the maximum material flux associated with the outflow is  
               approximately 5\% of the total accretion rate through the outer boundary. 
               Thus, $F_\mm{outflow}$ depends on the disk temperature; colder disks yield
               lower $F_\mm{outflow}$ and vice versa.                       
  } 
\end{figure}
\section{ Electron-positron versus electron-proton jets}
Since the flow in the TL is heated through magnetic reconnection, it is naturally to ask
 whether it is $e^-/e^+-$ or $e^-/p-$dominated plasma. \\
In view of the  solar flares, we may expect the plasma in the TL to accommodate 
different plasma-populations such as electron-positron, electron-proton in a 
see of high $\gamma-$rays.  The relativistic acceleration that particles experience during 
reconnection events in the vicinity of the BH give rise to different types of pair creation, 
in particular $\gamma-\gamma$,  $\gamma-e^{\mp}$,  $e^{\mp}-e^{\mp}$,
$\gamma-p$,  $e^{\mp}-p$ \cite{Svensson82, Svensson91, White89}. 
Among other effects, pair fraction depends crucially on two important
 parameters: the optical depth to scattering and on the so called compactness parameter.
The latter parameter is a fundamental scaling quantity for relativistic plasmas having
$\vartheta= kT_\mm{e}/m_\mm{e}c^2$ is of order unity.\\
Taking into account that the plasma in the innermost region of the disk is freely falling, and that
$\Mdot_\mm{d} = 2 H_\mm{d} r <\rho_\mm{d}> U$, we obtain that the optical depth to scattering in the 
innermost part of  disk is:
\beq
  \tau^\mm{d}_\mm{s}  = \DD{\kappa_s \Mdot_\mm{d}}{2 r U} 
= 18.86 (\DD{r}{r_\mm{LSO}})^{-1/2} (\DD{\Mdot}{\Mdot_\mm{Edd}}),
\eeq 
 where $r_\mm{LSO}$ is the last stable radius. Noting that $\rho_\mm{W} = \eps^2 \rho_\mm{d}$,
 and that the optical depth obeys a similar relation, i.e., 
  $\tau^\mm{W}_\mm{s} = \eps^2 \tau^\mm{d}_\mm{s}$, we may conclude that the TL is optically thin to
  scattering for most accretion rate typical for AGNs. 
   Further, since the flow in the TL is steady and two-temperature, our models fall in the left-down
   corner of the $\tau_\mm{s}-\Mdot$ diagram of in Fig. 10 of Esin (1999),  from which we conclude that 
   jets are formed in ion/proton-dominated plasma.
\section{Extraction of rotational energies from the BH and from the disk}

A spinning black hole is surrounded by the so called ergosphere, inside which
no static observer 
is possible; any frame of reference must be dragged by the spin of the black hole. 
If the plasma in the ergosphere  is threaded by an external and ordered magnetic field,
then the MF-lines must be dragged as well, 
thereby generating an electric field $\vec{E}= \vec{V}\times \vec{B}$. 
In the ergosphere, however, negative energy orbits are possible (i.e., the total
energy including the rest mass of the particle is negative). If a PMF threads
the event horizon, it can put particles on  negative energy orbits. These particles will be swallowed
by the BH, while other particles with positive energy will emerge that carry 
electromagnetic energy flux from the hole. These extracted energies from the hole
is used then to accelerate the plasma-particles relativistically, possibly forming $e^-/e^+-$dominated jets
\cite[see][]{Blandford76, Rees82, Begelman84}.
 The direction and magnitude of the electromagnetic flux  are described by 
the Poynting vector: 
\beq
 \vec{P} = -\DD{1}{4 \pi} E \times B = -\DD{1}{4 \pi} V \times B\times B.
\eeq
Using  spherical geometry,  the components of the Poynting flux read:

\beq
 \begin{array}{ll}
\vec{P}  = -\DD{1}{4 \pi} \{ & 
           (\Vt \Bpr \Bt - \Vpr \Bt^2 - \Vpr \Bpz^2  + \Vpz \Bpr \Bpz), \\
       & (\Vpr \Bpr \Bpz - \Vpz \Bpr^2 - \Vpz \Bt^2  -\Vt \Bt^2), \\
       & (\Vpz \Bpz \Bt - \Vt \Bpz \Bt - \Vt \Bpr^2  + \Vpr \Bpr \Bt)\}.
\end{array}
\eeq
In the case of a spinning BH surrounded by a plasma with zero poloidal motion,
i.e., $ V=\Vpr = \Vpz = 0$, a PMF  
of external origin in the vicinity of  the event horizon is predominantly radial,
or equivalently, it has a monopole-like topology. In this case the Poynting
flux reads:
\beq
 \vec{P}^\mm{BH} \approx   - \DD{1}{4 \pi} (\Vt \Bpr \Bt, 0, 0)
\eeq
The corresponding electromagnetic luminosity carried out off the black hole
reads:
\beq
  L^\mm{BH}  = \int{\vec{P}^\mm{BH} _\mm{r} \cdot dS} = \DD{1}{4} r_\mm{H}^2 \Vt \Bpr \Bt,
\eeq
where $r_\mm{H}$ is the event horizon.

Similarly, applying the same procedure to the plasma in the TL, neglecting 
the effects of vertical advection, and noting that $\Bp=\eps \Bt$, then the 
Poynting flux is predominantly radial: 
\beq
 \vec{P}^\mm{TL} \approx  \vec{P}^\mm{TL} _\mm{r}  =  - \DD{1}{4 \pi} (\Vpr \Bt^2, 0, 0).
\eeq
The corresponding  luminosity is obtained then by integrating the flux over the radial
 surface of the TL at $r=\rtr$:
\beq
  L^\mm{TL}  = \int{\vec{P}^\mm{TL} _\mm{r} \cdot dS} = \rtr H_\mm{W} \Vpr \Bt^2.
\eeq
Thus, the ratio of the above two luminosities yields:
\beq
 \DD{L^\mm{BZ}}{L^\mm{TL}} 
 = \DD{1}{4 \eps } (\DD{\Vt^\mm{BH}}{\Vpr^\mm{tr}})
               (\DD{\Bpr^\mm{BH}}{\Bt^\mm{tr}})
                 (\DD{\Bt^\mm{BH}}{\Bt^\mm{tr}})
                   (\DD{r_\mm{H}}{\rtr})^2.
\eeq
How large this ratio is, depends strongly on the location of the transition radius $\rtr$.
We now turn to estimate the transition radius $\rtr$.\\
Noting that at $\rtr$  the magnetic energy is in equipartition with the thermal energy of the
disk-plasma which, in terms of Eq. A.2, means that:  
\beq
\DD{1}{\rtr^2} = \DD{1}{\rho} \DD{\D}{\D r} \Bpz^2.
\eeq
Furthermore, noting that $\rbl$ is the critical radius where $\Bp$ change topology, and where the
corresponding magnetic energy is roughly in equipartition with rotational energy, we obtain from
Eq. A.2: 
\beq
\DD{1}{\eps^2}\DD{1}{\rbl^2} = \DD{1}{\rho} \DD{\D}{\D r} \Bpz^2.
\eeq
Taking into account the spatial variations of $\Bp$ and $\rho$ as described in Equations
47 and 50, and solving for X at $\rbl$, we find that:
\beq
X(r=\rbl) = \eps^{4/3} \approx \left\{ \begin{array}{lll}
0.046 & ~~~~~~ {\rm if}~\eps = 10^{-1} & (\lra \rtr \approx 20) \\
0.018 & ~~~~~~ {\rm if}~\eps = 5\times 10^{-2} & (\lra \rtr \approx 50) \\
0.002 & ~~~~~~ {\rm if}~\eps =  10^{-2} & (\lra \rtr \approx 500),
\end{array} \right.
\eeq
where $\rtr$ is given in $\rbl-$units.
Using Eq. 34,  and taking into account the spatial variations of
$U_\mm{W}$ and $\Bt$ in the TL, and assuming that the
central BH is rotating at maximum rate, we may obtain the following 
 inequality inequality:
\beq
\DD{L^\mm{BZ}}{L^\mm{TL}} \le \DD{1}{4 \eps}{Y^{3/4}} 
(\DD{r_\mm{H}}{\rtr})^2 =  \DD{1}{4 \eps} (\DD{r_\mm{H}}{\rtr})^{5/4},
\eeq
which is larger than $5\%$ for most reasonable values of $ \eps (> 10^{-3})$. \\
Consequently, the TL appears to provide the bulk of the electromagnetic energy associated 
 with jets, even when the  central BH is  maximally rotating. The contribution of 
 the BHs to the total electromagnetic energy flux is significant, but it
 is unlikely to be dominant.\\
Thus, the total power available from the launching layer and from the spinning BH is:
\beq
 L_\mm{tot} =  L_\mm{TL} +  L_\mm{BZ} =  L_\mm{TL}( 1 + \DD{1}{4\eps} (r_\mm{H}/\rtr)^{5/4}).
\eeq
If the accretion rate is relatively large and the BH rotates at maximum rate,
 $\tau_\mm{TAW}$,  becomes longer than $\tau_\mm{dyn}$, and therefore $\rtr$ must move toward smaller radii. 
Consequently, the ratio
 ${L^\mm{BZ}}/{L^\mm{TL}}$ increases, but remains below 25\% of the total available
luminosity from the TL for most reasonable values of  accretion rates.

We note that  $\rtr$ in our present consideration serves as a lower limit. 
The reason is that the rate of turbulence generation is likely to
starts decreasing when the ratio of the magnetic to internal energies 
exceeds the  saturation value that has been revealed by MHD calculations \cite{Hawley96}, and 
is likely to suffer of a complete suppression at $\rtr$, where $\beta$ starts to exceed unity.
\begin{figure*}
\begin{center}
{\hspace*{-0.5cm}
\includegraphics*[width=6.5cm]{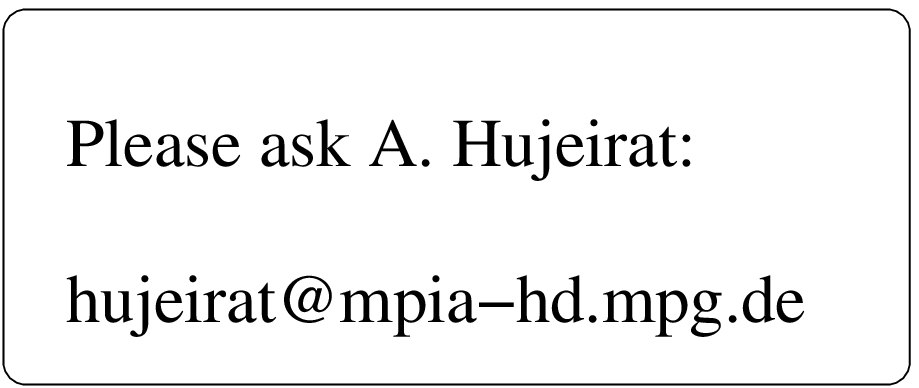}
}
\end{center}
{\vspace*{-0.4cm}}
\caption [ ] { The distribution of the velocity field in the transition region 
           superposed on equally-spaced isolines of the
          angular velocity $\Omega$. The strong-decrease of $\Omega$
          with radius in the equatorial region relative to its slow-decrease in the
          TL is obvious.  The overlying dynamically unstable corona is rotation-free. 
          The orange colour corresponds to large material densities,  violet to intermediate
          and blue correspond to extremely tenuous coronal plasma. 
          The 2D distributions of the variable have been obtained using a vertical interpolation
          procedure. 
  } 
\end{figure*}
\begin{figure*}
\begin{center}
{\hspace*{-0.5cm}
\includegraphics*[width=6.5cm]{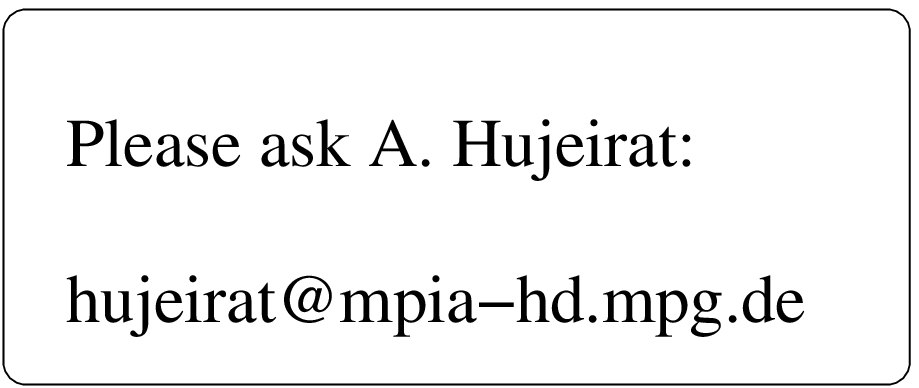}
}
\end{center}
{\vspace*{-0.4cm}}
\caption [ ] {20 equally-spaced isolines of the poloidal-component $ \rm B_\mathrm{P}$
 (white contours),  superposed on the coloured-distribution of the toroidal magnetic
            field. Blue, green and red colours correspond to low, intermediate and
             to high TMF-values, respectively.
  } 
\end{figure*}
\section{Luminosity of truncated disks} 

The primary source of heating in standard disks is due to dissipation of turbulence.
The corresponding heating function reads: \cite{Frank92}:
\beq
 D(r) = \DD{1}{2} \eta_\mm{tur} (r \DD{\D \Omega}{\D r})^2,
\eeq
where $\eta_\mm{tur} = \Sigma \nu_\mm{tur}$, $\Sigma $ is surface density,
 and $\nu_\mm{tur}$ is the standard turbulent viscosity.\\
Following the analysis of \cite{Frank92}, the r-integrated time-independent
equation of angular momentum in one-dimension reads:
\beq
\Sigma U \Omega =  \eta_\mm{tur} \DD{\D \Omega}{\D r} + C/2\pi r^3,
\eeq
where C is an integration constant. The classical assumption  was that $\D \Omega/ \D r$ must vanishes at some radius close
to the central object, and so imposing a physical constrain  on C.
The idea here was motivated by the two observationally supported assumptions:
1) Almost all astrophysical objects, that so far have been observed,  are found to 
   rotate at sub-Keplerian rates, and 2)
   $\nu_\mm{tur}$ is a non-vanishing function/process.\\
In the recent years, however, two possibilities have emerged:
1) several accretion disk in AGNs appear to truncate
   at radii   much larger than their gravitational radii, 2) 
 magnetic fields in accretion disks can be amplified  by the Balbus-Hawley instability,
 and potentially they may reach thermal equipartition,  beyond which 
 local random  motions,  turbulence-generation as well as  turbulence dissipation are suppressed.\\ 
Therefore, we may think of C as some sort of a magnetic effect that come into play
whenever the accretion flow approaches a certain radius $\rtr$ from outside, at which  turbulence
generation is diminished.
 In view of Equation 41, we obtain:
\beq
 \eta_\mm{tur} = \DD{\Mdot}{3\pi}[1 - (\DD{\rtr}{r})^{1/2}] \LRA 
 D(r) = \DD{3GM\Mdot}{8 \pi r^3}[1 - (\DD{\rtr}{r})^{1/2}],
\eeq
which applies for $r \ge \rtr$ only. The corresponding luminosity reads:
\beq
L = 2 \int^\mm{\rtr}_{\infty} D(r) 2\pi r dr=  \DD{1}{2} \DD{G M \Mdot}{\rtr}.
\eeq
In the case of the nucleus in M87, if the disk surrounding the BH truncates at 
100 gravitational radii, then the corresponding accretion luminosity is:
\beq
 L   \approx \DD{G M \Mdot}{2 R_{tr}} = \DD{G M \Mdot}{200\,R_{g}} 
  = 2.5\times 10^{-3} c^2 \Mdot = 2.5\times 10^{-6} c^2 \Mdot_\mm{Edd} \approx 10^{42} erg\, s^{-1},
\eeq
where $\Mdot = 1.6\times10^{-3}\Mdot_\mm{Edd}$ has been used \cite{DiMatteo03}.\\
Thus, the jet power should of the order of:
\beq
 L_\mm{J} = \DD{G M \Mdot}{2 r_{BL}} - \DD{G M \Mdot}{2 r_{BL}} \approx \DD{G M \Mdot}{2 R_{BL}} 
  \approx 10^{44} erg\,s^{-1},
\eeq
which agrees with the observations \cite[see][]{Bicknill99}.  In writing Eq. 45, we
have assumed that $\rbl$ is located very close to the last stable orbit.\\
Consequently, interior to $\rtr$ the energy goes primarily into powering the
jet with magnetic and rotational energies. This power dominate the luminosity
of the truncated disk, making it difficult to observe the disk directly.  
\section{Discussion}
In this paper we have presented a theoretical model for accretion-powered jets in 
AGNs and microquasars. 
The model relies primarily  on electromagnetic extraction of rotational energy
 from the disk-plasma and the formation of  a geometrically thin launching layer
 between the disk and the overlying corona.\\
 The model is based on self-similar solution for the 3D axi-symmetric radiative
 two-temperature MHD equations. The very basic assumption underlying this model
 is that standard disks should truncate energetically at a certain radius $\rtr$,
 below which $\Bp$ start to   increase inwards as $r^{-2}$, and eventually becomes
 in excess of thermal equipartition.
 Thus, the matter and the frozen-in PMF can be accreted by the BH as far as  
  TAWs are able to efficiently extract rotational energy normal to the disk.
 Such a flow-configuration can be maintained if the PMF change topology at
 $\rbl$, which must be associated with loss of magnetic flux, so to avoid complete 
 termination of accretion. Moreover, 
  unlike standard disks in which the accretion
  energy is liberated away, or advected inwards as entropy (ADAF solutions), the accretion
  energy here is converted mainly into magnetic energy that power the jet. In this
  case, the plasma in the innermost part of the disk becomes: 1) turbulent-free, 2) rotates
  sub-Keplerian,  and 3) remains relatively cold and confined to a geometrically thin disk.
  On the other hand,  the plasma in the TL
   rotates super-Keplerian, dissipative, two-temperature, 
     virial-hot and ion-dominated. \\
  To first order in $\eps$, the profiles of the main variables in the disk are:
\beq
   U_\mm{d}(r) =  [U^2_\mm{0,d} + \DD{2}{\eps^2}(\DD{1}{X} -1)]^{1/2},
\eeq
\beq
   \rho_\mm{d}(r) = \rho^0_\mm{d} {\cal F}_{\rho} X^{-3/2},
\eeq
\beq
   \Omega_\mm{d}(r) = \Omega^0_\mm{d} X^{-5/4},
\eeq
\beq
   T_\mm{d}(r) = T^0_\mm{d}  X^{-5/8},
\eeq
\beq
   B_\mm{p,d}(r) \approx B_\mm{\theta,d}(r) = B^0_\mm{d} X^{-2},
\eeq
where $X = r/\rtr$,
\({\cal F}_{\rho} = [(\Mdot^0_\mm{d} + 2 \Delta \Mdot_\mm{W})/\eps U_\mm{d}] \rtr^{-3/2}r^{-1/2}\)
(see Eq. 17),
$\rho^0_\mm{d} =\alpha^{-7/10}_\mm{tur} \Mdot^{11/20}r^{-15/8} f^{11/5}$,
\( \Omega^0_\mm{d} = \rtr^{-3/2}\) and  \(B^0_\mm{d}= \sqrt{\rho^0_\mm{d} T^0_\mm{d}}\).
Note that the superscript `0' and the subscript `d' denote the values at the equator and 
at $\rtr$, and which be found using standard accretion disk theory.  \\
 
In the launching region the variables  have the following profiles:
\beq
   U^\mm{W}(r) = \DD{\sqrt{2}}{\eps}  (1-\DD{1}{\sqrt{Y}})^{1/2} \rtr^{-1/4},
\eeq
\beq
   \rho^\mm{W}(r) = const. =
  \DD{\eps \Mdot^0_\mm{d}}{r^2 U_\mm{d} - 2\sqrt{2} \eps \Delta_0}|_{r=\rbl},
\eeq
\beq
   \Omega^\mm{W}(r) = \Omega^0_\mm{W} Y^{-5/4},
\eeq
\beq
   T^\mm{i}_\mm{W}(r) = T^\mm{i}_\mm{W,0}  Y^{-1/2},\,\,\,  T^\mm{e}_\mm{W}(r) = T^\mm{e}_\mm{W,0} Y^{-5/8}
\eeq
\beq
   B_\mm{p,W}(r)  =  B^0_\mm{p,W}  Y^{-2},
\eeq
\beq
  B_\mm{T,W}(r)  =  B^0_\mm{T,W}  Y^{-1/4},
\eeq
where $Y = r/\rbl$, \(\Delta_0 = [(r^{7/4}{\cal F}) -(r^{7/4}{\cal F})_{r=\rtr}]\), 
\(\Omega^0_\mm{W}= \rbl^{-3/2}\), \(T^\mm{i}_\mm{W,0} = 1/\eps^2\), \(T^\mm{e}_\mm{W,0} = 1/\eps\),
\( B^0_\mm{p,W} =  1/\eps^2\) and \( B^0_\mm{T,W} =  1/\eps^2\). 
 The variables here are given in  the units listed in Table A.1 (see Appendix). \\ 

 Furthermore, we would like to point out that:
\ben
\item   Magnetic reconnection in combination with Joule heating is an
  essential ingredient for our model to work.  The magnetic diffusivity adopted here,
  $\nu_\mm{mag} = H_\mm{W} \Vat$, is proportional to $\Vat$, and therefore it vanishes
  at turning points where
  $\Bt$ changes its sign. This prescription  does not smooth the $\Bt-$gradients 
  on the account of Joule-heating the plasma, but it strengthen the $\Bt-$gradients
  even more, thereby enhancing 
  the rate of $\Bt-$reconnection. The underlying idea here is not to smooth the 
  $\Bp-$gradients, but to terminate the propagation of the TAWs to higher
  latitudes, trapping therefore the angular momentum in the TL.
  In general, a normal magnetic diffusivity cannot hinder the propagation of TAWs into the corona. 
\item  If the axi-symmetric assumption is relaxed, there might be still ways
     to transport angular momentum from the disk into the corona without forming
     a dissipative TL. Such ways can be explored using a full 3D implicit radiative
     MHD solver, taking into account Bremsstrahlung and Synchrotron coolings, 
     heat conduction, and able to deal with highly stretched mesh distributions. 
     Such  solvers are not available to date, and therefore beyond the
     scope of the present paper.
     On the other hand, since the region of interest is located in the vicinity of
     an axi-symmetric black hole,  the axi-symmetric assumption might be a valid  
     description for the flow in the TL and in the disk.  
     Furthermore, in the absence of heating from below, coronae around  black holes have
     been found to be dynamically unstable to heat conduction \cite{Hujeirat02}.
     In order to hold a BH-corona against dynamical collapse, the MFs threading the 
     corona must be sufficiently strong. These, in turn, force the gyrating 
     electrons to cool rapidly through emitting Synchrotron radiation.  
     In the absence of efficient heating mechanisms for the plasma in the corona, 
     this cooling may even runaway. \\
     Consequently, coronae around BHs most likely will diminish, and the robust way
     to form jets would be through the formation of a TL, which 
     might be a natural consequence of accretion flows onto BHs. 
\item We note that if the innermost part of the disk is threaded by an $r^{-2}-$PMF, then
      $\Omega \sim r^{-5/4}$ is likely to be the optimal  profile for the angular velocity.
      Specifically, 
\ben 
\item $\Omega \sim r^{-5/4}$  satisfies Eq. 25, so that $\alpha_\mm{mag}$, and hence $\Mdot$
      can be fine-tuned so  to produce the radio luminosity and jet-power
      that  agrees with  observations.
\item Multi-dimensional radiative MHD calculations have shown that $r^{-5/4}$ is the 
      appropriate profile for the angular velocity in the TL \cite{Hujeirat03}.
\item Conservation of angular momentum implies that the flow in the disk must be
      advective-dominated to be able to supply the plasma in the TL with  
       angular momentum, so to maintain the $r^{-5/4}-$profile.
\item  The Keplerian profile $\Omega = c r^{-3/2}$, with c=1 yields slow accretion.
       In this case, the cooling time scales will be faster than the hydrodynamical time scale.
       Thus, the disk cools and the matter settles into a standard disk.  However, such a disk
       is unlikely to be stable if the PMFs  is in super-equipartition with the thermal
       energy. \\
       If $c<1$, then $\rtr$ must move to much larger radii, and as $\Bp$ increases
       inwards as $r^{-2}$, accretion may terminate by the magnetic pressure. 
\item Adopting the profile $\Omega\sim r^{-1}$ does not fulfil  
      Eq. 25, which requires that the bulk of the generated TMF-energy must be advected
       with the outflow.  Moreover, this profile yields unacceptably large and
      radius-independent toroidal magnetic field in the transition layer, and so
      overestimating the radio luminosity.
\een
\item  The transition radius $\rtr$, which is treated as input parameter,
             depends strongly on the accretion rate.
       If the outer disk fails to supply its innermost part with a constant
accretion rate, then the flow may become time-dependent. In this case, the emerging
plasmas does not show up  as continuous jet, but  rather in a blob-like configuration.
A high $\Mdot$  gives rise to $\tau_\mm{TAW} > \tau_\mm{hd}$ so that $\rtr$ must move towards
smaller radii.  In this case, the total rotational power extracted from the innermost
part of the disk via TAWs becomes smaller, giving rise thereby  to a less energetic and
weakly collimated jets. On the other hand, a low $\Mdot$ 
would allow $\rtr$ to move outwards, the total surface from which angular momentum
is extracted is larger, and the resulting jets are more energetic, have larger magnetic and
 rotational energies and therefore  are more collimated. 
\item  We have shown that the BZ-process modifies the total power of the jet.
        However, we have verified also that the  extraction of rotational energy from the innermost 
        part of the disk continue to dominate the total power available from the disk-BH system.
 \een
Finally, in a future work, we intend to implement our model to study the jet-disk connection
in the microquasar GRS 1915-105 in the radio galaxy M87.  

\appendix
\section{}
\begin{table}[htb]
\begin{tabular}{ll} 
Scaling variables &   \\\hline
Mass: &   $\tilde{\cal M} = 3\times 10^8M_{\odot}$   \\
Accretion rate: &   $\tilde{\Mdot} = 10^{-3}\Mdot_{Edd}$   \\
Distance: & $\tilde{R} = R_\mm{in} = 3R_\mm{S},$ where $R_\mm{S} = 2G\tilde{\cal M}/c^2$ \\
Temperature: &  $\tilde{\cal T} = 5 \times 10^7 K $  \\
Velocities: &  $\tilde{V} = \tilde{V_\mm{S}} = 
       [\gamma {\cal R}_\mm{gas} \tilde{\cal T}/\mu_\mm{i}]^{1/2},\, \mu_\mm{i} = 1.23 $ \\
Ang. Velocity: & $\tilde{\Vt} = \tilde{V_\mm{Kep}} = (G\tilde{\cal M}/\tilde{R})^{1/2}$, 
$\LRA \eps = \tilde{V_\mm{S}}/\tilde{\Vt} $ \\
Magnetic Fields: & $ \tilde{B} = \tilde{V_\mm{S}}/\sqrt{4 \pi \tilde{\rho}} $  \\
Density: &  $ \tilde{\rho} = \tilde{\Mdot}/(\tilde{H_\mm{d}} \tilde{R_\mm{out}} \tilde{V_\mm{S}})$ 
$ = 2.5 \times 10^{-12} g\,cm^{-3} $ \\\hline
\end{tabular}
\caption{The scaling variables used to reformulate the equations in non-dimensional form.}
\end{table}
We use spherical geometry to describe the inflow-outflow configuration around a Schwarzschild black hole.
The flow is assumed to be 3D axi-symmetric, and the two-temperature description is used to study
the energetics of the ion- and electron-plasmas.  \\
In the following we present the set of equations in non-dimensional form using the scaling
 variables listed in Table A.1.
\ben
\item The continuity equation:
\beq
   \DD{\D \rho}{\D t} + \DD{1}{r^2}\DD{\D}{\D r}{(r^2 \rho U)}
    +  \DD{1}{ r \cos{\theta}}\DD{\D}{\D \theta}{(\cos{\theta} \rho V)} =  0
\eeq
\item The radial momentum  equation:
    \[\DD{\D U}{\D t} + {U}\DD{\D U}{\D r}
    +   \DD{V}{r}\DD{\D U}{\D \theta}
    = \DD{1}{\rho }\DD{\D P}{\D r} + \DD{1}{\eps^2}(r \Omega^2 -\DD{1}{r^2}) \]
\beq
    + \DD{1}{\rho} \DD{\Bpz}{r} \DD{\D  }{\D \theta} \Bpr
    - \DD{1}{\rho} \DD{1}{r} \DD{\D  }{\D r} (r \Bpz^2)
    - \DD{1}{\rho} \DD{1}{r} \DD{\D }{\D r}  (r \Bt^2)
\eeq
\item The vertical momentum  equation:         
  \[ \DD{\D V}{\D t} + {U}\DD{\D V}{\D r}
    + \DD{V}{r \cos{\theta}}\DD{\D V}{\D \theta}
    = \DD{1}{r \rho}\DD{\D P}{\D \theta} - \DD{1}{\eps^2}\tan{\theta}\DD{\Vt^2}{r} \]
\beq
    - \DD{1}{\rho} \DD{\Bpr}{r} \DD{\D r \Bpz}{\D r}  
    - \DD{1}{2 r \rho}  \DD{\D}{\D \theta} (\Bpr^2 +\Bt^2)
\eeq
\item The angular momentum  equation: 
 \[  \DD{\D \ell }{\D t}
    + \DD{1}{r^2}\DD{\D}{\D r}{(r^2 \rho U \ell)}
    + \DD{1}{r \cos{\theta}}\DD{\D}{\D \theta}{(\cos{\theta} V \ell)}
    = \Bpz \DD{\D \Bt}{\D \theta}
    +  \Bpr \DD{\D r \Bpz}{\D r} \]
\beq
    +  \DD{\cos{\theta}^2}{r^2} \DD{\D}{\D r} ({r^4 \rho \nu_\mm{tur} \DD{\D \Omega}{\D r}})
    + \DD{1}{\cos{\theta}} \DD{\D}{\D \theta} ({ \cos{\theta}^3 \rho \nu_\mm{tur} \DD{\D \Omega}{\D \theta}})
\eeq
\item The  internal  equation of the ions:
\[   \DD{\D \vepsdi }{\D t}
    + \DD{1}{r^2}\DD{\D}{\D r}{(r^2 \rho U \vepsdi)}
    + \DD{1}{ r \cos{\theta}}\DD{\D}{\D \theta}{(\cos{\theta} V \vepsdi)}
    = -(\gamma -1) \vepsdi \nabla \cdot V \]
\beq
    +  (\gamma -1) \Phi - \Lambda_\mm{i-e}  
    + \nabla \cdot \kappa^\mathrm{cond}_\mathrm{i} \nabla \Ti
\eeq
\item The  internal  equation of the electrons:
\[
   \DD{\D \vepsde }{\D t}
    + \DD{1}{r^2}\DD{\D}{\D r}{(r^2 \rho U \vepsde)}
    + \DD{1}{ r \cos{\theta}}\DD{\D}{\D \theta}{(\cos{\theta} V \vepsde)}
    = -(\gamma -1) \vepsde \nabla \cdot V  \]
\beq
    +  (\gamma -1) \Phi + \Lambda_\mm{i-e} - \Lambda_\mm{\rm B }
    -  \Lambda_\mm{\rm C } - \Lambda_\mm{\rm Syn }
    + \nabla \cdot \kappa^\mathrm{cond}_\mathrm{e} \nabla \Te
\eeq
\item The equation of the radial component of the poloidal magnetic field:
\beq
   \DD{\D \Bpr}{\D t}
    + \DD{1}{r^2 \cos{\theta}} \DD{\D \veps_\mathrm{emf}}{\D \theta} =0
\eeq
\item The equation of vertical component of the poloidal magnetic field:
\beq
   \DD{\D \Bpz}{\D t}
    -   \DD{1}{r \cos{\theta}} \DD{\D \veps_\mathrm{emf} }{\D r} =0
\eeq
\item The equation of toroidal magnetic field:
 \[  \DD{\D \Bt }{\D t}
    + \DD{\eps}{r}\DD{\D}{\D r}{(r U \Bt)}
    + \DD{1}{r}\DD{\D}{\D \theta}{ (V \Bt)}
    = \DD{1}{\eps} r \cos{\theta} \Bpr \DD{\D \Omega}{\D r}
    +  \DD{1}{\eps}\cos{\theta} \Bpz \DD{\D \Omega}{\D \theta} \]
\beq
    + \DD{\eps^2}{r} \DD{\D}{\D r} (\nu_\mm{mag} \DD{\D}{\D r}{r \Bt})
    + \DD{1}{r^2} \DD{\D}{\D \theta} (\nu_\mm{mag} \DD{\D}{\D \theta}{\Bt})
\eeq
\een
where the subscripts ``i'' and ``e'' correspond to ions and electrons.
$\rho,~\vec{V}=(V_r,V_\theta,V_\varphi)$ and P   are the density, velocity vector,
  and the  
gas pressure $P (\doteq  {\cal R}_{\rm gas} \rho (T_i/\mu_{\rm i} + T_e/\mu_{\rm e})),$ respectively.
$\vepsdi,\,\vepsde$ denote  
the internal energy densities due to ions and electrons,
${\cal E}^{e,i}= P^{e,i}/(\gamma-1),$ where $\gamma=5/3$, $\mu_{\rm i}=1.23$ and $ \mu_{\rm e}=1.14 $.
$\vec{B}=(B_{r},B_{\theta},\Bt)=(\Bp,\Bt)$ is the magnetic field, and
$\veps_\mathrm{emf}$ is the modified electromotive force taking into account the contribution
of the magnetic diffusivity.  
$\Phi$, $\Lambda_{B}$, $\Lambda_{\rm i-e}$, $\Lambda_{\rm C}$, $\Lambda_{\rm syn}$ are
the turbulent dissipation rate, Bremsstrahlung cooling, Coulomb coupling between the ions
and electrons, Compton and synchrotron coolings, respectively. These processes read:
 \[ \Phi = \nu_\mm{mag} |\nabla \times B|^2/{\cal N}\]
 \[\Lambda_{\rm i-e} = 5.94\times 10^{-3} n_i n_e c 
      k \DD{(T_{\rm i} - T_{\rm e})}{T^{3/2}_{\rm e}} /{\cal N}\]
\[ \Lambda_{\rm B} =  4ac \kappa_\mathrm{abs} \rho (T^4 - E)/{\cal N}, \]
\[ \Lambda_{\rm C} = 4 \sigma n_e c  (\DD{k}{m_e c^2}) (T_e - T_{rad})E/{\cal N}, \]
where $\kappa_\mathrm{abs}$ and $\sigma$ are the absorption and scattering  coefficients.
${\cal N} = [(\gamma-1)/\gamma] (\tilde{V}^2 \tilde{\Vt}/\tilde{R})$ is a normalization
quantity (see Table A.1).
$n_{\rm e},~n_{\rm i}$ are the electron- and ion-number densities. 
E is the density of the radiative energy, i.e., the zero-moment of the radiative field.
The radiative temperature  is defined as $T_{rad}=E^{1/4}$.
For describing synchrotron cooling $\Lambda_\mm{ Syn}$, Eq. 24 is used.
\end{document}